\documentclass[11pt,letterpaper]{article}
\title{A CARTOGRAPHY FOR 2x2 SYMMETRIC GAMES}
 \usepackage[latin1]{inputenc}
 \usepackage{amsmath}
 \usepackage{graphics}
 \usepackage{chicago}
 \usepackage{amssymb,color,graphpap}
 \usepackage{ae,graphics,enumerate}
 
\begin{document}
 {\center
 \begin{minipage}[c][22cm][c]{14cm}
  \makeatletter

  \makeatother
  \begin{center}
   \vfill
   \thispagestyle{empty}
   {\scriptsize III CONGRESO COLOMBIANO Y I CONFERENCIA ANDINA INTERNACIONAL DE INVESTIGACI\'ON DE OPERACIONES}\vspace{1cm}

   {\large A CARTOGRAPHY FOR 2x2 SYMMETRIC GAMES}\vspace{1cm}

   {\large UNA CARTOGRAF\'IA PARA LOS JUEGOS SIM\'ETRICOS DE 2x2}\vspace{1cm}

   \'Alvaro Francisco Huertas-Rosero\vspace{1cm}

   Universidad de los Andes, Bogot\'a, Colombia

   ahuertas@uniandes.edu.co\vspace{1cm}
 
   RESUMEN\vspace{11pt}

   \parbox{14cm}{
    Se propone una representaci\'on bidimensional del espacio de juegos sim\'etricos de 2x2 en su representaci\'on estrat\'egica.   Esta representaci\'on provee una herramienta para la clasificaci\'on de los juegos sim\'etricos de 2x2, cuantificaci\'on de la fracci\'on que tiene una cierta caracter\'istica, y predicciones sobre los cambios en las caracter\'isticas de un juego cuando se realizan cambios en la matriz de ganancia que lo define.}\vspace{1cm}

   ABSTRACT\vspace{11pt}

  \parbox{14cm}{
   A bidimensional representation of the space of 2x2 Symmetric Games in the strategic representation is proposed.   This representation provides a tool for the classification of 2x2 symmetric games, quantification of the fraction of them having a certain feature, and predictions of changes in the characteristics of a game when a change in done on the payoff matrix that defines it.}\vspace{1cm}

   PALABRAS CLAVE

   2x2 games, Strategic Representation, Classification of Games\vspace{1cm}
  \end{center}
 \end{minipage}}
 \bibliographystyle{chicago}
 \section{INTRODUCTION}
  There is a large amount of investigation on dynamic models based on 2x2 symmetric nonzero-sum games \citeN{Evolution}, \citeN{Selection}, where a certain payoff, and some times some uncertainty on it \citeN{Subjective} is chosen to construct the model.   The features of the static games defined by the payoff matrices are very important in these dynamic models \citeN{Suitable}, but so far we lack a complete cartography of the space of possible payoff matrices to guide our selection for a certain model.
 
 \section{SYMMETRIC 2x2 GAMES IN THE STRATEGIC FORM}
  Among the nonzero sum games, the most simple kind of game is probably the 2x2 symmetric with imperfect information.  And the most simple way to represent it is the strategic form introduced in \citeN{TheoryGamesEB}.\\
 \begin{equation}
  \Game = (\{A,B\},\{0,1\} \otimes \{0,1\},\$_{A}:\{0,1\}\otimes\{0,1\}\to \mathbb{R},\$_{B}:\{0,1\}\otimes\{0,1\}\to \mathbb{R})
\end{equation}
  where $\Game$ represents the game, $\{A,B\}$ is the set of players, $\{0,1\}$ is the space of strategies of any of the players, and the composite space $\{0,1\}\otimes\{0,1\}$ is the space of \textit{positions}.  $\$_A$ and $\$_B$ are the payoff function for the two players. This is a function that goes from the space of positions to some ordered set, which is generally $\mathbb{Z}$, the set of the integer numbers.  These payoff functions are sometimes referred as \textit{payoff matrices}, taking the labels of the strategies as indexes
  \begin{equation}
   (\mathbb{P}_A)_{ij} =  \$_A(i,j)
  \end{equation}
  In symmetric games, there is a symmetry condition on the payoffs, which ensures equality of conditions for the two players \citeN{Simetricos}.  If we interchange players, both the indexes $i$, $j$ are interchanged, and the payoff matrices $\mathbb{P}_{A}$, $\mathbb{P}_{B}$ are interchanged.
  \begin{equation}
   \mathbb{P}_{A}=(\mathbb{P}_{B})^t
  \end{equation}

 \section{ESSENTIAL AND UNESSENTIAL FEATURES IN THE PAYOFF BIMATRIX}
  An acceptable payoff can be defined in any set with an order relation \citeN{OrdinalUtility}, but it proves to be very useful to define it within a compact set.   If we assign a numerical payoff to both players in every choice situation, we need 8 numbers.  The condition of symmetry reduces them to 4.  But as it was stated in \citeN{LinearInvariance} and \citeN{LinearUtility}, the properties of the game cannot depend either on the value of payoff represented by number 0 (additive invariance), or an overall factor in the scale of payoffs shared by both players (scale invariance).  There are, then, two parameters that can be ruled out, and we are left with only \textbf{two numbers}.

  Let's represent the payoff matrices for A and B in the following way:
  \begin{equation} \label{Gclasica}
   \mathbb{P}_{A}=
    \begin{pmatrix}a & b \\ c & d \end{pmatrix}\hspace{1cm}
   \mathbb{P}_{B}=
    \begin{pmatrix}a & c \\ b & d \end{pmatrix}
   \end{equation}
  The mean of this numbers is irrelevant (additive invariance), so can substract it from the four parameters (that is $a'=a-\frac{1}{4}(a+b+c+d)$) and the properties of the game are preserved.

  We can also use the scale invariance to take rid of the overall scale considering (a',b',c',d') as a vector in $\mathbb{R}^4$ and normalizing. ($a'' = a'/norm(a',b',c',d')$).

  To make the two relevant parameters explicit, we choose a SO(4) transformation wich takes $a''$ into $\frac{1}{2}(a''+b''+c''+d'')$ wich we know is always zero, and get, for example,these new parameters:
  {\small \begin{equation}\label{transformacion}
   (G_0,G_A,G_B,G_{AB})=(a,b,c,d)
   \frac{1}{2}\begin{pmatrix}
    1 & 1 & 1 & 1\\
    1 & 1 & -1 & -1\\
    1 & -1 & 1 & -1\\
    1 & -1 & -1 & 1
    \end{pmatrix}
   \end{equation}}
   As it was stated above, $G_{0}$ is allways zero, but $G_{A}$ is the difference between the expected payoffs of each player if \textit{the other} player chooses 0 y 1 with equal probability.  $G_{B}$ is the difference in \textit{the other player}'s payoffs if each player plays an uniform distribution of probabilities 0 and 1.   $G_{AB}$ is the payoff difference between the two symmetric situations (A choosing the same as B)

 \section{GEOMETRIC REPRESENTATION}
   Normalization made the vector $(G_{A},G_{B},G_{AB})$ unitary, and therefore we have a representation of the space of relevant parameters as the surface of a sphere.

   The conditions for the existence of Nash equilibria and Pareto Optimal conditions are sets of inequalities involving these three parameters.  Each inequality defines a plane in the parameter state that cuts the sphere in a certain way.

   Every possible game in the considered type, defined by a payoff matrix, has a ``normalized'' representative in the surface of the sphere (except for the trivial one with the same payoff for any situation).  \textbf{The fraction of surface of the unit sphere enclosed by the planes corresponding to the conditions is exactly the fraction of the possible games set wich fulfills those conditions}.

 \section{NASH EQUILIBRIA (NE)}
  The definition for a Nash Equilibrium (NE from now on) is simple: if both player decrease their payoff departing individually from a certain choice condition, then this condition is a NE.\citeN{Nash Equilibrium}

  For example, the position($i^*$,$j^*$) is a NE iff:
  \begin{equation}
   \forall (k,l)\epsilon \{positions\}\hspace{16pt}  \mathbb{P}_{i^*j^*} \geqslant \mathbb{P}_{kj^*}\hspace{16pt}and\hspace{16pt}
   \mathbb{P}_{i^*j^*} \geqslant \mathbb{P}_{i^*l}
  \end{equation}
  In a symmetric game, the existence of a pure strategy NE is guaranteed \citeN{NashSym}, and this makes things easier.  A Pure strategy Nash Equilibrium will correspond to a maximum payoff in a column of player A's payoff matrix and a maximum payoff in a row of player B's payoff matrix.  In terms of the parameters defined in \ref{transformacion}:

   \begin{equation}\label{NashConditions} \begin{aligned}
    (-1)^i(G_{A}+ (-1)^j G_{AB}) &\geqslant 0\\
    (-1)^j(G_{A}+ (-1)^i G_{AB}) &\geqslant 0
   \end{aligned} \end{equation}
   Where the indexes $i$ and $j$ can take values within the set of strategies $\{0,1\}$.   A remarkable feature of these conditions is that they only involve the parameters $G_{A}$ and $G_{AB}$.  They are completely insensitive to $G_{B}$, the average difference in payoff owed to a change in the \textit{other player} strategy. 
   \subsection{Symmetric positions}
    For the symmetric positions (i=j) it is easily seen that both conditions are the same: 
    \begin{equation}
    (-1)^i(G_{A}+(-1)^i G_{AB}) \geqslant 0
    \end{equation}
   This condition will be fulfilled for half the possible games, for it defines a plane that cuts the sphere in two halves, both for $(1-i)(1-j)=1$ and $ij=1$.  The planes defined by the condition ``00 is NE'' and ``11 is NE'' are orthogonal, and that allows us to infer that $\frac{1}{4}$ of the games will have only (0,0) as a Nash equilibrium,  $\frac{1}{4}$ of the games will have (1,1) as a Nash equilibrium and $\frac{1}{4}$ of the games will have \textit{both positions} as Nash equilibria.
   \subsection{Nonsymmetric positions}
    For nonsymmetric positions 01 and 10 the conditions \ref{NashConditions} are different ($s_1=1, s_2=-s_3$) from each other, and define, again, perpendicular planes:
    \begin{equation}\begin{aligned}
    (-1)^iG_{A}-G_{AB} &\geqslant 0\\
    (-1)^jG_{A}-G_{AB} &\geqslant 0
    \end{aligned}\end{equation}
   However, the conditions are identical for position 01 and for position 10. This means that \textbf{If one nonsymmetric position is NE, the other is NE too}.
  \subsection{ZONES IN THE G SPACE}
\begin{figure}[htb]
    \center
    \includegraphics{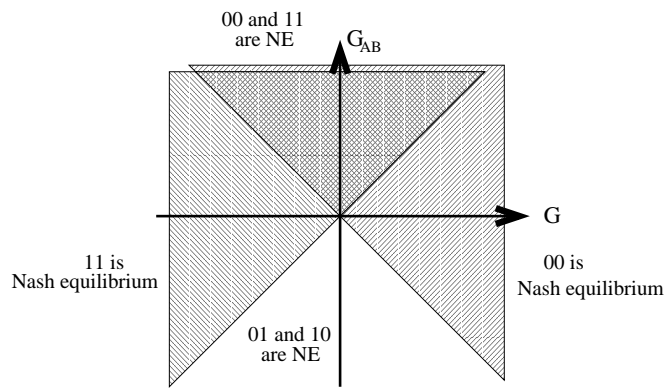}
    \caption{Zones with different NE in the $G$ parameters  space}\label{Nash2d}
   \end{figure}

   In figure \ref{Nash2d} three zones in the space of arameters $G_{AB}$ and $G_A$ are shown that are separated by the Nash Equilibrium Conditions.
 \section{PARETO OPTIMAL POSITIONS (PO)}
  A position is Pareto-optimal (PO from now on) when each player can improve his payoff, but doing so, he diminishes the other player's payoff \citeN{Pareto}.

  One of the conditions for a certain choice situation to be a PO is clearly that \textbf{it is not a NE}. The other, concerning the other player's payoff maximization, can be viewed as the NE condition on the \textit{transposed} payoff matrix.  Transposing payoff matrix amounts to interchanging $G_{A}$ and $G_{B}$, so we already have them:
   \begin{equation}\label{ParetoConditions} \begin{aligned}
    s_1 G_{B}+ s_2 G_{AB} &\geqslant 0\\
    s_1 G_{B}+ s_3 G_{AB} &\geqslant 0
   \end{aligned} \end{equation}

  \subsection{The geometric complete picture}
   NE conditions generates planes defined by $(G_{A}\pm G_{AB}=0)$, and OP conditions generate planes defined by $(G_{B}\pm G_{AB}=0)$.  To achieve some symmetry in our partition of the sphere we need to cut the sphere with the plane $(G_{A}\pm G_{B}=0)$ as well.  This involves comparing one diagonal element of the payoff matrix with the other diagonal element, and comparing one nondiagonal element with the other nondiagonal element.
   \begin{figure}[hbt]
    \center
    \includegraphics{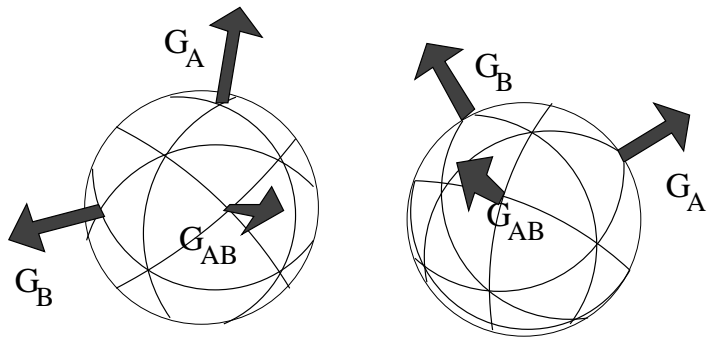}
    \caption{Partition of the unit sphere}\label{PartSphere}
   \end{figure}

  We cut the sphere with three pairs of orthogonal planes, getting 24 identical regions, which will allow us to compute easily the fraction of the sphere corresponding to certain set of such NE or PO conditions. Every region of the sphere characterized by NE and OP conditions will be conformed by some of these 24 small curve ``triangles'', which we will call \textit{elementary regions} from now on They are shown in figure \ref{PartSphere}.

    If we choose the greatest absolute value as the norm instead of the euclidean norm, the space of possible games is projected into the \textbf{surface of the unit cube}.

\begin{figure}[htb]
    \center
    \includegraphics{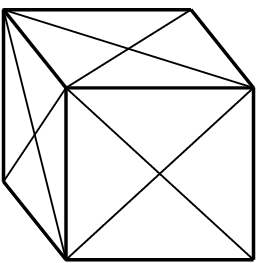}\hspace{1cm}
    \includegraphics{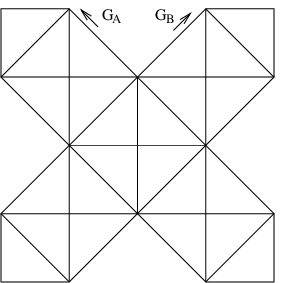}
    \caption{Partition of the unit cube, and its unfolded form}\label{PartCube}
   \end{figure}

    The Nash and Pareto conditions will cut this cube in a very symmetrical way, as they did to the sphere.     It becomes apparent then that all the elementary regions are alike.  In figure \ref{PartCube} this partition of the cube is shown, together with the unfolded surface as it is going to be presented later.   In the central square $G_{AB}>0$, while in the square formed by the four extreme triangles of the unfolded cube $G_{AB}<0$
  \section{A classification of 2x2 nonzero sum symmetric games}
   We can see the EN conditions  affecting $G_{AB}$ and $G_{A}$ and the PO conditions affecting $G_{AB}$ and $G_{B}$ as classification criteria, meaning the following for a certain position:
   \begin{itemize}
    \item If the position is NE it means that each player gets a maximum payoff \textbf{for himself}.
    \item If the position is  PO it means that each player produces a maximum payoff \textbf{for the other player}.
   \end{itemize}
    The importance of the first criteria is evident by itself, but the other criteria becomes important in cooperative game \citeN{Threats} where players are allow to interact before choosing, or in dynamic games, where in some cases players learn that it can be advantageous to behave ``generously'' \citeN{Behaviour}.

    Normally the PO definition include the inverse of the NE condition (the PO payoff must be susceptible to be raised by any player) \citeN{Pareto}.  But here we are going to try a less restrictive approach, taking the conditions on the transpose of the payoff matrices as \textit{independent} from the conditions of the actual payoff matrix.

    Using these NE and PO conditions independently we can assign each region of the sphere a certain kind of game .  The NE and PO planes cut the sphere into 12 regions, with 3 different sizes. There are 8 triangular (elementary) regions covering each $\frac{1}{24}$ of the sphere, 4 bigger triangular regions covering each $\frac{1}{12}$ of the sphere (conformed by two elementary regions), and  2 square regions covering each $\frac{1}{6}$ of the sphere, conformed by 4 elementary regions.

    The labeling of strategies 0 and 1, on the other hand, is arbitrary, and only diagonal (symmetric) and nondiagonal (nonsymmetric) positions are distinguishable (if the game is symmetric, this is a natural distintion).  Then we would have 9 kinds of games, the same result obtained by \citeN{Rapoport0} and \citeN{Robinson} for symmetric games

    It can be wise, however, to discriminate wether a certain diagonal P0 position (pure or mixed) have higher or lower payoff than a certain NE position.  This give us 12 kinds of games.

  \subsection{ORDER GRAPHS}
   In a taxonomy of 2x2 games \citeN{Rapoport0}, a very useful graph was proposed, where all the possible outcomes of payoffs of the players are plotted, payoff of A vs. payoff of B.

    From these an order graph can be constructed that allows to find Nash Equilibrium and Pareto optimal positions, as is shown in \citeN{Robinson}.   It can therefore be used as a classification scheme for strictly-ordered-payoff games (where there are no two equal elements in the payoff matrix).

   To construct an order graph, the four points corresponding to the four positions are connected with arrows when a player can unilaterally shift from one to other, and the arrows point in the direction where the payoff is higher for the involved player.

    If A's payoff is plotted in the Y axis, then A's arrows point allways upwards and B's arrows points allways to the right.  Doing so, we get an \textbf{order graph}. In this order graph,\textbf{a Nash Equilibrium is a point where all the adjacent arrows point at}.

   To find PO we draw arrows that point in the direction where the \textbf{other player}'s payof is increased.   We will call these arrows \textit{Pareto arrows}.  The points fulfilling PC are the points where the Pareto arrows converge.Conversely, we will call the arrows ponting in the direction where the own payoff increases \textit{Nash arrows}.

\begin{figure}[hbt]
     \center
     \includegraphics{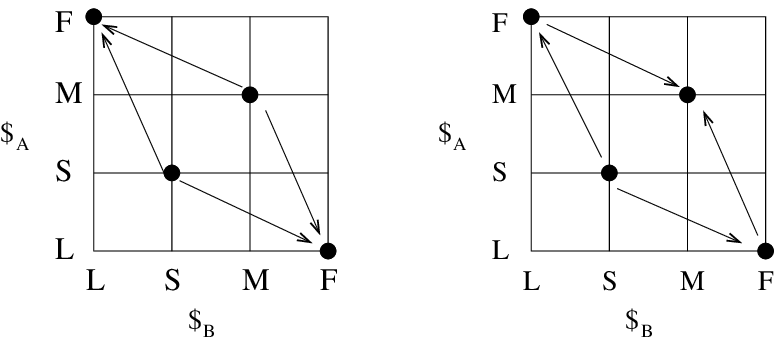}
     \caption{Order Graph for the Prisoner's dilemma with Nash arrows and Pareto arrows}\label{Robinson0}
    \end{figure}

   As an example, the payoff matrix for the Prisoner's Dilemma is given by:
   \begin{equation}
    \begin{matrix}
     \ & \begin{matrix} \text{\scriptsize B confesses} & \text{\scriptsize B acuses A} \end{matrix}\\
    \begin{matrix} \\ \text{\scriptsize A confesses} \\ \text{\scriptsize A acuses B} \end{matrix} &
     \begin{pmatrix}
      \text{\scriptsize Moderate sentence} & \text{\scriptsize Severee sentence}\\
      \text{\scriptsize Freedom} & \text{\scriptsize Lifetime sentence}
     \end{pmatrix}
    \end{matrix}
   \end{equation}
   In figure \ref{Robinson0} the order graph for this game is presented, where L=lifetime, S=severe, M=moderate and F=freedom. 

    If we transpose the payoff matrix, the Nash arrows turn into Pareto arrows, and we get a different game. When we transpose, the axis are interchanged in the graphs, and only \textbf{arrows with negative slope} change their direction.  All the information we need to characterize a game is in the Nash arrows.  Moreover, if we mark the NE, we can infer the directions of the arrows, and the game is completely determined.

\begin{figure}[bht]
   \center
   \includegraphics{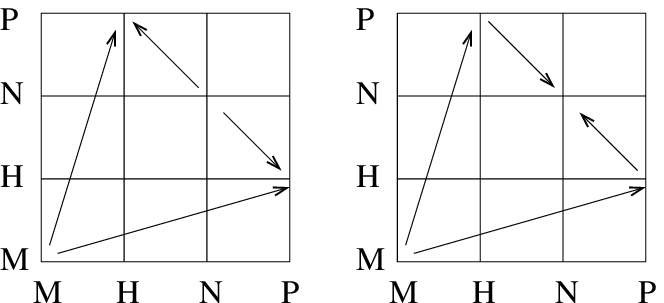}
   \parbox{1cm}{$\longrightarrow$\vspace{1.5cm}\\}
   \includegraphics{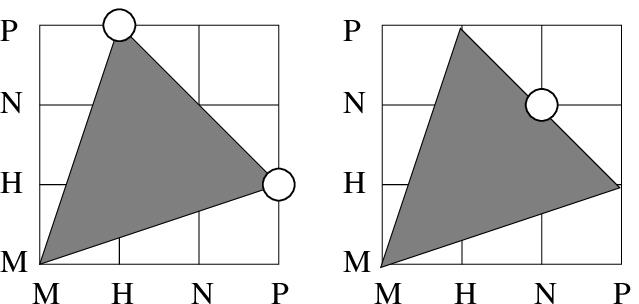}
   \caption{Simplified Robinson Diagrams}\label{SimpliRobinson}
  \end{figure}

    For example:  Let's consider a game with payoff matrix $\begin{pmatrix}4 & 5 \\ 1 & 0 \end{pmatrix}$ similar to the ``Chicken Game''.  In figure \ref{SimpliRobinson} we show the Robinson graphs for the game and its transposed-matrix game, and a simplified version of both graphs.

  \section{A LIST OF 2x2 NONZERO-SUM SYMMETRIC GAMES}
  \renewcommand{\thefootnote}{\fnsymbol{footnote}}
  In this work, a classification for these games is proposed where, as a first criterion, the diagonal and non-diagonal NE are counted, resulting in three categories:
  \begin{itemize}
   \item Games with one diagonal NE (table \ref{OneDiag})
   \item Games with two diagonal NE (table \ref{TwoDiag})
   \item Games with two non-diagonal NE (table \ref{TwoNond})
  \end{itemize}

   \begin{table}[hbt]
    \center \small
    \begin{tabular}{|ccccp{1cm}p{4cm}|}\hline
     PO & $\$_{NE(diag)}>\$_{PO}?$ & \# Triangles & Fraction & {\scriptsize Diagram} & examples\\\hline\hline
     is NE & --- & 4 & {\normalsize $\frac{1}{6}$} & \parbox{1cm}{\includegraphics{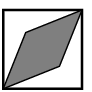}} & 
     \parbox{2.8cm}{ \vspace*{3pt}\textbf{Cholesterol: friend or foe}$^*$\\
       \citeN{GTCourse}}
     \parbox{1cm}{\tiny $\begin{pmatrix}4 & 2 \\ 3 & 1 \end{pmatrix}$} \vspace{3pt}\\\hline
     not NE &  Yes & 2 & $\frac{1}{12}$ & \parbox{1cm}{\includegraphics{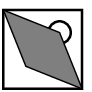}} & 
     \parbox{2.8cm}{ \vspace*{3pt}\textbf{Hershey's kisses} \citeN{Kisses},\\
     \textbf{Deadlock}\\ \citeN{Miracle}}
     \parbox{1cm}{\tiny $\begin{pmatrix}3 & 4 \\ 1 & 2 \end{pmatrix}$}\\\hline
     not NE & No & 2 & $\frac{1}{12}$ & \parbox{1cm}{\includegraphics{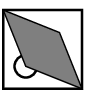}} & 
     \parbox{2.8cm}{\vspace*{3pt}\textbf{Prisoner's Dilemma} \\ \citeN{SymGames}}
     \parbox{1cm}{\tiny $\begin{pmatrix}3 & 1 \\ 4 & 2 \end{pmatrix}$}\\\hline
     Two$^\dagger$  & Yes & 3 & $\frac{1}{8}$ & \parbox{1cm}{\includegraphics{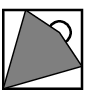}\\\includegraphics{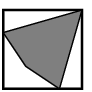}} & 
     \parbox{1cm}{\tiny $\begin{pmatrix}4 & 5 \\ 2 & 1 \end{pmatrix}$\\
      $\begin{pmatrix}5 & 4 \\ 1 & 2 \end{pmatrix}$ }\\\hline
     Two$^\dagger$ & No & 1 & $\frac{1}{24}$ & \parbox{1cm}{\includegraphics{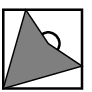}} & 
     \parbox{1cm}{\tiny $\begin{pmatrix}3 & 5 \\ 2 & 1 \end{pmatrix}$}\\\hline
    \end{tabular}
    \caption{Games with one diagonal NE}\label{OneDiag}
   \end{table}

   Games with only one NE (table \ref{OneDiag}) are maybe the most studies among 2x2 symmetric games.   That is the reason for having three examples in table \ref{OneDiag} with an own name.

      \footnote[1]{\textbf{Cholesterol: friend or foe} is a game where the expected payoff matrix depends of a probability, and can change between two extreme scenarios.  The game referred corresponds to one of those scenarios.}
      \footnote[2]{If \textbf{two PO} exist in these games, they can be both diagonal and  non-diagonal.  In the first case, the PO position to be compared with the NE is the one that is not a NE, and in the second case what we compare is a \textit{mixed strategies} PO.}

   Games with two diagonal NE (table \ref{TwoDiag}) have only one interesting feature: the equilibrium choice dilemma.  There is, moreover, one only kind of games within this category.

   \begin{table}[htb]
    \center \small
    \begin{tabular}{|cccp{1.1cm}p{4cm}|}\hline
     PO &  \# Triangles & Fraction & {\scriptsize Diagram} & Example\\\hline\hline
     Both NE & 6 & {\normalsize $\frac{1}{4}$} & 
      \parbox{1.1cm}{\includegraphics{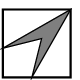}\\\includegraphics{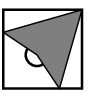}\\\includegraphics{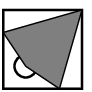}} &
      \parbox{2.8cm}{\vspace*{3pt}\textbf{Pareto Coordination Game}$^*$\\
       \citeN{GamesTV}}
      \parbox{1.1cm}{\tiny $\begin{pmatrix}4 & 1 \\ 2 & 3 \end{pmatrix}$\\
       $\begin{pmatrix}5 & 1 \\ 4 & 3 \end{pmatrix}$\\
       $\begin{pmatrix}5 & 1 \\ 3 & 2 \end{pmatrix}$\\
        } \vspace{3pt}\\\hline
    \end{tabular}
    \caption{Games with two diagonal NE}\label{TwoDiag}
   \end{table}
   \begin{table}[hbt]
    \center \small
    \begin{tabular}{|ccccp{1cm}p{4cm}|}\hline
     PO & $\$_{NE(mixed)}>\$_{PO}?$ & \# Triangles & Fraction & {\scriptsize Diagram} & examples\\\hline\hline
     One & Yes & 1 & {\normalsize $\frac{1}{24}$} & \parbox{1cm}{\includegraphics{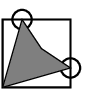}} & 
     \parbox{1cm}{\tiny $\begin{pmatrix}3 & 2 \\ 5 & 1 \end{pmatrix}$} \vspace{3pt}\\\hline
     One &  No & 1 & $\frac{1}{24}$ & \parbox{1cm}{\includegraphics{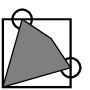}} & 
     \parbox{2.8cm}{ \vspace*{3pt}\textbf{Chicken Game} \citeN{Kisses}}
     \parbox{1cm}{\tiny $\begin{pmatrix}4 & 2 \\ 5 & 1 \end{pmatrix}$}\\\hline
     Two & --- & 4 & $\frac{1}{6}$ & \parbox{1cm}{\includegraphics{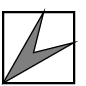}} &
     \parbox{1cm}{\tiny $\begin{pmatrix}2 & 3 \\ 4 & 1 \end{pmatrix}$}\\\hline
    \end{tabular}
    \caption{Games with two non-diagonal NE}\label{TwoNond}
   \end{table}

   Games with two nondiagonal equilibria (table \ref{TwoNond}) are interesting mainly because the equilibrium choice dilemma is solvable by using \textit{mixed strategies}. These games have a stable mixed strategy NE.  In this category we find the famous ``Chicken Game''

  \section{CONCLUSIONS}
\begin{figure}[htb]
     \center
     \includegraphics{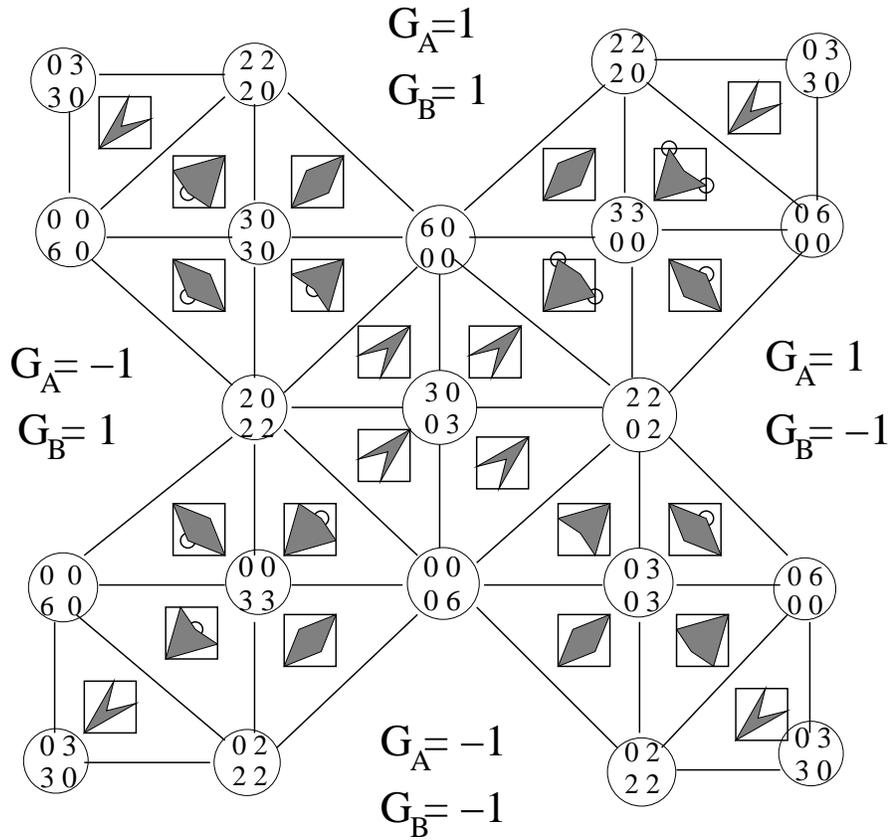}
     \caption{Map of all possible 2x2 symmetric games}\label{Map}
    \end{figure}

   The whole classification of games proposed here can be represented in a map of the games as that shown in figure \ref{Map}.  This is a scheme of all triangular regions of the unfolded unit cube where the $G$ parameters of the payoff matrices can be projected.

   To help the using of the map, some ``canonical'' payoff matrices are shown in the vertexes of the triangles.  Any payoff matrix can thus be decomposed in a trivial part (a matrix with all its entries identical) and a convex combination of the canonical matrices in the vertexes of the corresponding triangle.

   The point in the unit cube representing a certain payoff matrix wil be nearer to the vertex whose canonical matrix has a larger quoefficient in the decomposition.   The distance from each vertex is inversely proportional to the decomposition quoefficient, even though there is an irrelevant overall scale.

   Among the uses we can find for the map, the following are the most obvious:
   \begin{itemize}
    \item Compute the fraction of games having a certain feature.
    \item Check which changes in a payoff matrix can change the features of the game, and how such features change.
    \item Find easily a payoff matrix to define a game with certain characteristics.
   \end{itemize}
  
  A good example to ilustrate the use of the map in figure \ref{Map} is the game ``Cholesterol: friend or foe''.   This game have a payoff matrix that is a convex combination of two payoff matrices weighted by probabilities:
  \begin{equation}
   \mathbb{P} = 
    \begin{pmatrix} 9 & 15 \\ 5 & 7 \end{pmatrix}\  P\  + 
    \begin{pmatrix} -9 & -3 \\ -1 & 1 \end{pmatrix}\  (1-P) 
  \end{equation}
  We can decompose each matrix in the relevant and irrelevant part, and write the former in terms of the ``canonical'' payoff matrices shown in the map.  We obtain:
   \begin{figure}[hbt]
    \center
    \includegraphics{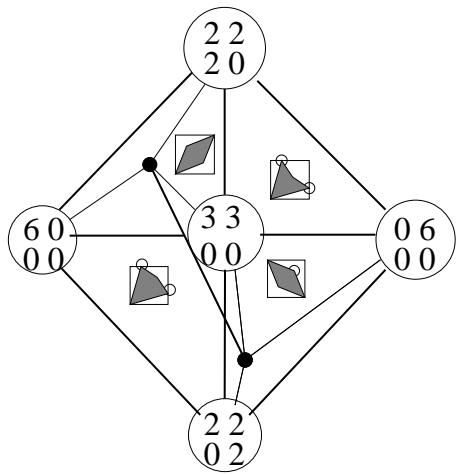}
    \caption{Position of ``Cholesterol: frien or foe'' for different p}\label{Trajectory}
   \end{figure}
    \begin{equation}
     \begin{pmatrix}9 & 15 \\ 5 & 7\end{pmatrix} =
     \frac{16}{6}\left(\frac{3}{8}\begin{pmatrix}0 & 6 \\ 0 & 0\end{pmatrix}
     + \frac{3}{8}\begin{pmatrix}2 & 2 \\ 0 & 2\end{pmatrix}
     + \frac{2}{8}\begin{pmatrix}3 & 3 \\ 0 & 0\end{pmatrix} \right)
     + \begin{pmatrix}5 & 5 \\ 5 & 5\end{pmatrix}
    \end{equation}
   For cholesterol being healthy and
    \begin{equation}
     \begin{pmatrix}-9 & -3 \\ -1 & 1\end{pmatrix} =
     \frac{6}{24}\left(\frac{2}{24}\begin{pmatrix}0 & 0 \\ 0 & 6\end{pmatrix}
     + \frac{4}{24}\begin{pmatrix}0 & 0 \\ 3 & 3\end{pmatrix}
     + \frac{18}{24}\begin{pmatrix}0 & 2 \\ 2 & 2\end{pmatrix}\right)
     + \begin{pmatrix}-9 & -9 \\ -9 & -9 \end{pmatrix}
    \end{equation}
   For cholesterol being unhealthy.

   Spanning the possible values of the probability $P$ we move along the map as is shown in figure \ref{Trajectory}

 \bibliography{JuegosCuanticos}

\begin{thebibliography}{}

\bibitem[\protect\citeauthoryear{Dehejla}{Dehejla}{2003}]{Kisses}
Dehejla, R. (2003).
\newblock Notes of class for ``microeconomics and policy analysis''.
\newblock http:/www.columbia.edu/itc/sipa/u8216-03/PDF/f00lab6.pdf/.

\bibitem[\protect\citeauthoryear{Flitney and Abbot}{Flitney and
  Abbot}{2003}]{Miracle}
Flitney, A.~P. and D.~Abbot (2003, October).
\newblock Advantage of a quantum player over a classical one in 2x2 quantum
  games.
\newblock {\em PROCEEDINGS OF THE ROYAL SOCIETY OF LONDON SERIES A-MATHEMATICAL
  PHYSICAL AND ENGINEERING SCIENCES\/}~{\em 459\/}(2038), 2463--2474.
\newblock http://www.arxiv.org/abs/quant-ph/0209121.

\bibitem[\protect\citeauthoryear{GameTheory.net}{GameTheory.net}{2003}]{GamesT%
V}
GameTheory.net (2003).
\newblock Gametheorynet1.
\newblock http://www.gametheory.net/html/popular.html.

\bibitem[\protect\citeauthoryear{Harsanyi}{Harsanyi}{1980}]{LinearInvariance}
Harsanyi, J.~C. (1980).
\newblock {\em A General Theory of Equilibrium Selection in Games}.
\newblock MIT press, Cambridge, Massachusets.

\bibitem[\protect\citeauthoryear{Heap and Varoufakis}{Heap and
  Varoufakis}{1995}]{OrdinalUtility}
Heap, S. P.~H. and Y.~Varoufakis (1995).
\newblock {\em Game Theory: A Critical Introduction}.
\newblock Routledge, London.

\bibitem[\protect\citeauthoryear{Lewontin}{Lewontin}{1961}]{Evolution}
Lewontin, R.~C. (1961).
\newblock Evolution and the theory of games.
\newblock {\em J. Theor. Biol.\/}~{\em 1}, 382--403.

\bibitem[\protect\citeauthoryear{Luce and Raiffa}{Luce and
  Raiffa}{1965}]{LinearUtility}
Luce, R.~D. and H.~Raiffa (1965).
\newblock {\em Games and Decisions}.
\newblock John Wiley, New York.

\bibitem[\protect\citeauthoryear{Marshall}{Marshall}{2003}]{Suitable}
Marshall, J. A.~R. (2003, June).
\newblock On the suitability of the 2 x 2 games for studying reciprocal
  cooperation and kin selection.
\newblock eprint: arXiv:cs.GT/0306128 v2.

\bibitem[\protect\citeauthoryear{Matsushima}{Matsushima}{1998}]{Subjective}
Matsushima, H. (1998).
\newblock In {\em Towards a Theory of Subjective Games}, Number CIRJE-F-9 in
  CIRJE F-Series. Faculty of Economics, University of Tokyo.
\newblock http://www.e.u-tokyo.ac.jp/cirje/research/dp/98/cf9/contents.htm.

\bibitem[\protect\citeauthoryear{Morris}{Morris}{1994}]{SymGames}
Morris, P. (1994).
\newblock {\em Introduction to Game Theory}.
\newblock Springer, N.Y.

\bibitem[\protect\citeauthoryear{Owen}{Owen}{1968a}]{NashSym}
Owen, G. (1968a).
\newblock {\em Game Theory}.
\newblock W.B.Saunders, New York.

\bibitem[\protect\citeauthoryear{Owen}{Owen}{1968b}]{Pareto}
Owen, G. (1968b).
\newblock {\em Game Theory}.
\newblock W.B.Saunders, New York.

\bibitem[\protect\citeauthoryear{Owen}{Owen}{1968c}]{Threats}
Owen, G. (1968c).
\newblock {\em Game Theory}.
\newblock W.B.Saunders, New York.

\bibitem[\protect\citeauthoryear{Possajennikov}{Possajennikov}{2002}]{Behaviou%
r}
Possajennikov, A. (2002, January).
\newblock In {\em Cooperative Prisoners and Aggressive Chickens: Evolution of
  Strategies and Preferences in 2x2 games}, Number 02-04 in
  SonderForschungBereich 504. University of Mannheim.
\newblock http://smealsearch.psu.edu/18937.html.

\bibitem[\protect\citeauthoryear{Rapoport and Guyer}{Rapoport and
  Guyer}{1978}]{Rapoport0}
Rapoport, A. and M.~Guyer (1978).
\newblock A taxonomy of 2 x 2 games.
\newblock {\em General Systems\/}~{\em 11}, 203--214.
\newblock Cited in: http://www.sysc.pdx.edu/classes/552w01w.htm.

\bibitem[\protect\citeauthoryear{Ratliff}{Ratliff}{2003}]{GTCourse}
Ratliff, J. (2003).
\newblock Graduate-level course in game theory.
\newblock http://virtualperfection.com/gametheory/.

\bibitem[\protect\citeauthoryear{Robinson}{Robinson}{1998}]{Robinson}
Robinson, D. (1998).
\newblock {\em Notes of class for the course on Strategic Thinking}, Chapter
  Classifying Symmetric 2X2 Games.
\newblock Laurentian University, Canada.
\newblock http://www.economics.laurentian.ca.

\bibitem[\protect\citeauthoryear{Vaughan}{Vaughan}{1995}]{Selection}
Vaughan, R. (1995, August).
\newblock Evolutive equilibrium selection i: Symmetric two-player binary choice
  games.
\newblock ESRC working papers. The ESRC Centre for Economic Learning and Social
  Evolution.
\newblock http://else.econ.ucl.ac.uk/papers/ees.pdf.

\bibitem[\protect\citeauthoryear{von Neumann and Morgenstern}{von Neumann and
  Morgenstern}{1944}]{TheoryGamesEB}
von Neumann, J. and O.~Morgenstern (1944).
\newblock {\em The Theory of Games and Economic Behavior}.
\newblock Princeton University Press.

\bibitem[\protect\citeauthoryear{Weibull}{Weibull}{1996}]{Simetricos}
Weibull, J.~W. (1996).
\newblock {\em Evolutionary Game Theory}, pp.\  25 to 31.
\newblock MIT press, Cambridge, Mass.

\end{thebibliography}
\end{document}